\documentclass[pra,reprint,amsmath,amssymb,showkeys,graphics,graphicx,showkeys]{revtex4-1}

\usepackage{times,graphicx,graphics,amsmath,amssymb,bm,xcolor}
\usepackage[colorlinks,linkcolor=red,citecolor=blue,urlcolor=brown]{hyperref}

\newcommand{\ket}[1]{\ensuremath{\vert #1 \rangle}}

\newcommand{\be}{\begin{equation}}
\newcommand{\ee}{\end{equation}}
\newcommand{\ben}{\begin{eqnarray}}
\newcommand{\een}{\end{eqnarray}}
\newcommand{\bes}{\begin{subequations}}
\newcommand{\ees}{\end{subequations}}
\newcommand{\bF}{\begin{figure}}
\newcommand{\eF}{\end{figure}}

\newcommand{\avg}[1]{\langle #1 \rangle}
\def\ket#1{ | #1 \rangle}

\def\tr{ {\rm{Tr }}\,}

\newcommand{\arxiv}[2][arxiv:]{\href{http://arxiv.org/abs/#1#2}{#1#2}}

\begin{document}

\title{Quantum discord as a resource in quantum communication}

\author{Vaibhav Madhok}
\email{vmadhok@gmail.com}
\affiliation{Department of Physics and Computer Science, Wilfrid Laurier University, Waterloo, Ontario N2L 3C5, Canada}
\affiliation{Center for Quantum Information and Control, University of New Mexico, Albuquerque, NM 87131-0001, USA}

\author{Animesh Datta}
\email{animesh.datta@physics.ox.ac.uk}
\affiliation{Clarendon Laboratory, Department of Physics, University of Oxford, OX1 3PU, United Kingdom}

\date{\today}

\begin{abstract}
As quantum technologies move from the issues of principle to those of practice, it is important to understand the limitations on attaining tangible quantum advantages. In the realm of quantum communication, quantum discord captures the damaging effects of a decoherent environment. This is a consequence of quantum discord quantifying the advantage of quantum coherence in quantum communication. This establishes quantum discord as a resource for quantum communication processes. We discuss this progress, which derives a quantitative relation between the yield of the fully quantum Slepian-Wolf protocol in the presence of noise and the quantum discord of the state involved. The significance of quantum discord in noisy versions of teleportation, super-dense coding, entanglement distillation and quantum state merging are discussed. These results lead to open questions regarding the tradeoff between quantum entanglement and discord in choosing the optimal quantum states for attaining palpable quantum advantages in noisy quantum protocols.
\end{abstract}


\maketitle

\section{Introduction}

Quantum information science has shown that the devices and protocols governed by the laws of quantum mechanics can have information processing capabilities superior to their classical counterparts. It has shed light on what properties of quantum systems could be harnessed for building future technology and engineering applications. The role of  information, quantified via measures of correlations such as entanglement between sub-systems, provides vital clues to the superior information processing capabilities of devices based on quantum mechanics. As emphasized by Landauer\cite{l61,Berut12}, information is physical and indicates that the information processing capabilities of a device are not independent of the physics that governs its operation.

At a more fundamental level, understanding how physical systems process and exchange information is crucial to gaining insights into the workings of our universe. For example, the connections between entropy, information and thermodynamics form the cornerstone of statistical mechanics. A relevant  example is Feynman's path-integral formulation of non-relativistic quantum mechanics.  In this approach, physical phenomena are described by \emph{events}. An event can occur through various alternatives or paths, each of which is characterized by a complex probability amplitude, that has both real and imaginary components in general. If these paths are in principle indistinguishable, i.e. there is no information whatsoever in the universe that can help us distinguish them, then the corresponding probability amplitudes add up causing interference. A canonical instance is the two-slit experiment. As long as there is no information available as to which slit the photon takes, we see interference fringes on the screen. The photon is considered to be in a superposition of two wave packets, centered around the classical path out of each slit. The question as to why the universe cares about which ``path" the photon takes is both philosophical and intriguing. It certainly tells us that the universe cares about certain kinds of information or the lack of it in formulating its laws.

Quantum information science has added a whole new perspective to the study of quantum mechanics. This has resulted in a better understanding of quantum phenomena like the entanglement and decoherence, and given us the tools to view certain quantum properties of physical systems as a resource.
In this spirit, this article is devoted to the study of the nature of quantum correlations themselves in the light of quantum information. Quantum correlations are believed to be at the heart of the weirdness of quantum mechanics since the days of Einstein\cite{epr}, and the resource for the potential benefits quantum information processing might provide. Computational algorithms and communication protocols based on quantum mechanics that accomplish tasks more efficiently than the best known classical methods are scenarios where such benefits can be reaped. Instances include quantum algorithms for integer factorization\cite{ekert96a,shor} and searching an unsorted database\cite{grov}. The advantages in quantum cryptography enable communication with guaranteed security against eavesdropping\cite{grtz02}.

One of the simplest yet most intriguing primitives in quantum information theory is quantum teleportation, or entanglement-assisted teleportation\cite{nc00}. It is the process by which an unknown quantum state can be transmitted from one location to another, without the state being transmitted through the intervening space. Expressing teleportation as a resource inequality
\be
\label{teleport}
[qq]  + 2 [c \rightarrow c] \succeq  [q \rightarrow q],
\ee
shows that a shared ebit and 2 bits of classical communication to communicate a single \textit{unknown} quantum bit. Here, we introduce the notation used in the resource theory of quantum communication protocols\cite{dhw08}. $[q \rightarrow q]$ represents one qubit of communication between two parties and $[qq]$ represents one shared ebit between two parties. Similarly, $[c \rightarrow c]$ represents one classical bit of communication between the parties. To communicate an unknown quantum bit by a classical procedure will take exponentially large amount of resources as compared to teleportation.

Another simple protocol which shows the advantage of quantum communications is the super-dense coding\cite{nc00}. Expressing it as a resource inequality\cite{dhw08},
\be
\label{superdense}
 [qq]  +  [q \rightarrow q] \succeq  2[c \rightarrow c],
\ee
showing that one can employ a shared ebit and a single bit of quantum communication to communicate 2 bits of classical information. Thus, complete classical information about two particles can be sent by direct manipulation of just one particle by the sender. The success of both these protocols rely on pre-existing shared entanglement. Without entanglement, it is impossible to execute either teleportation and the super-dense coding. However, quantum entanglement does not fully capture the quantum character of a system. There are several other possible resources to which quantum advantages are often ascribed. These include
\begin{enumerate}
\item size of Hilbert Space: The dimension of a quantum system of $n$ $d$-dimensional particles scales as $n^d.$ This is a consequence of the tensor product structure of quantum mechanics.

\item superpositions: A quantum state can exist in an arbitrary complex linear combination of classical logic states.    A classic example is the ``cat state", named after Schr\"{o}dinger.
\begin{equation}
|\psi_{cat} \rangle = \frac{|\mathrm{Alive} \rangle + | \mathrm{Dead} \rangle}{\sqrt{2}}.
\end{equation}
In a quantum process, both the `basis' states evolve in parallel according to a given unitary evolution.

\item interference: The quantum wave functions undergo interference, and different paths are explored in parallel in search of the solution and the probability amplitude of the path leading to the right solution gradually builds up.

\item indistinguishability of quantum states: Non-orthogonal quantum states cannot be unambiguously distinguished. Moreover, obtaining information about an unknown quantum state can cause disturbance and actually change it. This feature is exploited in designing cryptographic protocols.
\end{enumerate}

Yet, these do not comprise the whole story. Although entanglement is still generally believed to the resource of choice, in recent years there has been some progress in quantifying the quantum character of composite quantum systems using measures that go beyond entanglement. Quantum discord has been suggested as a prospective candidate and aims to captures all the quantum correlations in a quantum state\cite{ollivier01a}. There is a considerable interest in the research community about quantum discord, following evidence that it is responsible for the exponential speed up of a certain class of quantum algorithms over classical ones\cite{kl98,datta08a}.

An important question is whether quantum discord is merely a mathematical construct or does it have a definable physical role in information processing. It is known that there is a link between quantum discord and an actual physical task involving quantum communication between two parties -- an operational interpretation of quantum discord based on the quantum state merging protocol\cite{md10,cabmpw10}. Quantum discord is the markup in the cost of quantum communication in the process of quantum state merging, if one discards relevant prior information. A subsequent question regards the role of quantum discord in quantum information theory as a whole beyond the state merging protocol. We provide the answer to this question within the domain of quantum communication. For details on the properties of quantum discord, and its role in quantum computation, the reader is invited to several recent review articles\cite{celeri11,ds11,modi12}.

The key insight to the findings discussed here is that quantum measurements and environmental decoherence disturb a quantum system in a way that is unique to quantum theory. Quantum correlations in a bipartite system are precisely the ones that are destroyed by such disturbances, and therefore quantum communication protocols become overloaded by an amount exactly equal to quantum discord. More specifically, discord is the markup in the cost of quantum communication in the process of quantum state merging\cite{how05,how07}, if the system undergoes measurement and/or decoherence. We observe that quantum state merging protocol is a derivative of the more general fully quantum Slepian-Wolf (FQSW) protocol\cite{adhw09} and the closely related ``mother" protocol. A link between quantum discord and state merging can be generalised to a connection between quantum discord and the mother protocol and role of discord in essentially all bipartite, unidirectional and memoryless quantum communication protocols.

This is made possible by comparing the performance of the fully quantum Slepian-Wolf protocol in the presence and absence of decoherence and linking it to the discord of the state involved. While decoherence is expected to diminish the gain provided by a quantum protocol, we provide, for the first time, a general lower bound on the amount of this deterioration. Our bound is only dependent on the state involved, independent of the details of the protocol as well as the nature of the decoherence. Within the resource framework of quantum Shannon theory\cite{dhw04,dhw08}, we couple the performance of the FQSW protocol to the most general environmental decoherence to show that quantum discord of the state participating in the protocol is the lower bound to the depreciation of the protocol's performance. The FQSW protocol - a quantum communication-assisted entanglement distillation protocol -  is the parent protocol from which all information processing protocols emanate\cite{adhw09}. This protocol is the most general known in the family of protocols in quantum information theory. The significance of quantum discord in noisy versions of teleportation, super-dense coding, entanglement distillation and quantum state merging are discussed. We also demonstrate similar roles for quantum discord in quantum computation and correlation erasure. Our work shows that quantum discord captures and quantifies the advantage of quantum coherence in quantum communication. The generality of the FQSW protocol allows us to establish the role of quantum discord in the performance of noisy versions of quantum teleportation, super-dense coding, and distillation\cite{md11}.

Although it is known that entanglement is often necessary for the success of quantum protocols, and that the presence of decoherence affects its performance, we have now provided a quantitative result of the amount of such a depreciation. We show that the amount by which a protocol suffers in the presence of decoherence is an inherent property of the quantum states involved. It suggests that the choice of the best state for any noisy quantum protocol must be a tradeoff between the entanglement and discord of the state involved.  Given the non-monotonic relation between quantum discord and entanglement in quantum states\cite{modi10,alqasimi11,girolami11}, choosing the optimal state for a quantum task is a non-trivial one, though for which we now have the proper certificate.

\section{Classical and quantum communication}

Shannon provided the mathematical theory of communication, laying the foundation of classical information theory\cite{tc91}. During the latter half of the previous century, this led to enormous progress in understanding the key resources and issues surrounding communication and information technology. The central concept in classical information theory is the Shannon entropy, a measure of the uncertainty associated with a random variable. It quantifies the expected value of the information contained in a message, usually in the unit of bits. A `message' is a specific realization of the random variable, whose symbols or alphabets appear with probabilities $\{p(1), p(2), ..., p(i)...\}.$ The Shannon entropy $H$ associated with such a probability distribution is
\begin{equation}
     H= - \sum_{i} p(i) \log p(i).
\label{Eq:HKCT}
\end{equation}
This entropy is related to the physical resources required to solve certain information processing tasks. For example, Shannon's source coding theorem says the entropy represents an absolute limit on the best possible lossless compression of any source of information. This \textit{operational} interpretation of Shannon entropy in terms of lossless data compression is the cornerstone of classical information theory. Classical information theory also sheds light on the nature of correlations between two information sources, or, the input and the output of a channel and how they determine the rate at which messages can be exchanged securely. Shannon's noisy channel coding theorem states that the capacity of a channel is given by the maximum of the mutual information between the input and output of the channel, where the maximization is with respect to the input distribution. Another example pertinent to us is the Slepian-Wolf theorem\cite{sw73}. For that, consider a party Bob having access to some incomplete information about a random variable $Y,$ and another party Alice having the missing part $X.$ If Bob wishes to learn $X$ fully, how much information must Alice send to him ? Evidently, she can send $H(X)$ bits to satisfy Bob. However, Slepian and Wolf showed that she can do better, by merely sending $H(X|Y) = H(X,Y)-H(Y)$ bits, the conditional information\cite{tc91}.  Since $H(X|Y) \leq H(X)$, Alice can take advantage of correlations between $X$ and $Y$ to reduce the communication cost needed to accomplish the given task.

Quantum states can also be used for information processing and communication. In such cases, questions related to channel capacities, existence of a reliable compression scheme etc. about those quantum states become relevant. For example, Schumacher's quantum source coding theorem\cite{nc00} says that if $R \geq S(\rho),$ then there exists a reliable compression scheme of rate $R$ for an independent and identically distributed source of $\rho$. If $R \leq S(\rho)$ then no compression scheme of rate $R$ is reliable. Here $S(\rho)$  is the von Neumann entropy of the quantum state $\rho$. Another example which is useful to us is the quantum state merging protocol which is the extension of the classical Slepian-Wolf protocol into the quantum domain where Alice and Bob share the quantum state $\rho_{AB}^{\otimes n}$, with each party having the marginal density operators $\rho_{A}^{\otimes n}$ and $\rho_{B}^{\otimes n}$ respectively. Let $\ket{\Psi_{ABC}}$ be a purification of $\rho_{AB}.$ Assume, without loss of generality, that Bob holds $C.$ The quantum state merging protocol quantifies the minimum amount of quantum information which Alice must send to Bob so that he ends up with a state arbitrarily close to $\ket{\Psi}_{B'BC}^{\otimes n},$ $B'$ being a register at Bob's end to store the qubits received from Alice. It was shown that in the limit of $n\rightarrow \infty$, and asymptotically vanishing errors, the answer is given by the quantum conditional entropy\cite{how05,how07} $S(A|B) = S(A,B) - S(B)$. When $S(A|B)$ is negative, Bob obtains the full state with just local operations and classical communication, and distill $-S(A|B)$ ebits with Alice, which can be used to transfer additional quantum information in the future.

After realizing that quantum states can be used for information theoretic tasks, the next question is whether the conceptual breakthroughs promised by quantum communication can be palpably harnessed. In other words, what features of quantum mechanics can be used to give us information processing capabilities and communication protocols that are far superior to their classical counterparts.  We discuss both these questions in the next section in terms of quantum discord.

\section{Quantum Discord}
\label{sec:QD}

Characterizing the resources behind the enhancements and speedups provided by quantum mechanics over best known classical procedures is one of the most fundamental questions in quantum information science. Quantum entanglement\cite{hhhh07} is generally seen to be the key resource that gives quantum information processors their power. There are, however, quantum processes which provide an exponential advantage in the presence of little or no entanglement\cite{datta05a,datta07a}. In the realm of mixed-state quantum computation, for example, quantum discord\cite{ollivier01a,henderson01a} has been proposed as a resource\cite{datta08a} and there has been progress in this direction since\cite{e10,dvb10,ds11,fcco12}. It has also been shown to be a resource in quantum state discrimination\cite{rra11,lfwf11} and quantum locking\cite{dg09,boixo11}.

Quantum discord aims at generalizing the notion of quantum correlations in a quantum state, beyond entanglement\cite{ollivier01a,henderson01a}. It aims to capture all the nonclassical correlations in a quantum system. Quantum measurements disturb a quantum system in a way that is unique to quantum theory. Quantum correlations in a bipartite system are precisely the ones that are destroyed by such disturbances. As we discuss below, this feature of quantum systems can be used to quantify the amount of purely quantum correlations present in a bipartite quantum system.

Quantum mutual information is generally taken to be the measure of total correlations, classical and quantum, in a quantum state. For two systems, $A$ and $B$, it is defined as $ I(A:B) = S(A) + S(B) -S(A,B).$ Here $S(\cdot)$ denotes the von Neumann entropy of the appropriate distribution. For a classical probability distribution, Bayes' rule leads to an equivalent definition of the mutual information as $I(A:B) = S(A)-S(A|B).$  This motivates a definition of classical correlation in a quantum state.

Suppose Alice and Bob share a quantum state $\rho_{AB} \in \mathcal{H}_A\otimes \mathcal{H}_B.$ If Bob performs a measurement specified by the POVM  set $\{\Pi_i\},$ the resulting state is given by the shared ensemble $\{p_i,\rho_{A|i}\},$ where
$$
 \rho_{A|i} = \tr_{\!\!B}(\Pi_i\rho_{AB})/p_i,\;\;\;p_i=\tr_{\!\!A,B}(\Pi_i\rho_{AB}).
$$
A quantum analogue of the conditional entropy can then be defined as $\tilde{S}_{\{\Pi_i\}}(A|B)\equiv\sum_ip_iS(\rho_{A|i}),$ and an alternative version of the quantum mutual information can now be defined as $\mathcal{J}_{\{\Pi_i\}}(\rho_{AB}) = S(\rho_A)-\tilde{S}_{\{\Pi_i\}}(A|B),$ where $S(\cdot)$ denotes the von Neumann entropy of the relevant state. The above quantity depends on the chosen set of measurements $\{\Pi_i\}.$ To capture all the classical correlations present in $\rho_{AB},$ we maximize $\mathcal{J}_{\{\Pi_i\}}(\rho_{AB})$ over all $\{\Pi_i\},$ arriving at a measurement independent quantity
\be
\label{eq:J}
\mathcal{J}(\rho_{AB}) = \max_{\{\Pi_i\}}(S(\rho_A)-\tilde{S}_{\{\Pi_i\}}(A|B)).
\ee
Then, quantum discord is defined as\cite{ollivier01a}
\ben
\label{discexp}
\mathcal{D}(\rho_{AB}) &=& I(\rho_{AB})-\mathcal{J}(\rho_{AB}) \nonumber\\
                 &=& S(\rho_B)-S(\rho_{AB})+\min_{\{\Pi_i\}}\tilde{S}_{\{\Pi_i\}}(A|B)  \nonumber\\
                 &=& \min_{\{\Pi_i\}}\tilde{S}_{\{\Pi_i\}}(A|B) - S(A|B).
\een
The minimization can be restricted to rank-1 operators by supposing a POVM on $B$ can be fine-grained into
    $$
    \Pi_i = \sum_k \Pi_{ik}.
    $$
Then
 $$
p_{ik}\rho_{A|ik} =\tr_{B}(\rho_{AB}\Pi_{ik}),\;\;\;p_{ik} = \tr(\rho_{AB}\Pi_{ik}).
 $$
Evidently, $\sum_k p_{ik}=p_i$ whereby we can define $ p_{k|i} = p_{ik}/p_i. $ Also,
 \ben
\rho_{A|i} &=& \tr_B(\rho_{AB}\Pi_i)/p_i = \sum_k \frac{p_{ik}}{p_i}\tr_B(\rho_{AB}\Pi_{ik})/p_{ik} \nonumber \\
            &=& \sum_k p_{k|i}\rho_{A|ik},
 \een
and,
 \ben
 \sum_j p_j S(\rho_{A|j})&=&\sum_i p_i S\left(\sum_k p_{k|i}\rho_{A|ik}\right)\nonumber\\
                        &\geq& \sum_{i,k} p_i p_{k|i} S(\rho_{A|ik}) \nonumber\\
                        &=& \sum_{i,k}p_{ik} S(\rho_{A|ik}).
 \een
Since any POVM element can be written in terms of its eigendecomposition, the minimum conditional entropy, and therefore the discord is always attained on a rank-1 POVM.

For information theoretic considerations, the asymptotic limit needs to be studied. When Alice and Bob share $n$ copies of the state $\rho_{AB},$ we can define a regularized version of quantum discord as
\ben
\overline{\mathcal{D}}(\rho_{AB})&=&\lim_{n\rightarrow \infty} \frac{\mathcal{D}(\rho^{\otimes n}_{AB})}{n} \\ \nonumber
                                 &\equiv & I(\rho_{AB}) - \overline{\mathcal{J}}(\rho_{AB}),
\een
where
\be
\overline{\mathcal{J}}(\rho_{AB}) = \lim_{n\rightarrow \infty} \frac{\mathcal{J}(\rho^{\otimes n}_{AB})}{n}.
\ee
The quantity $\overline{\mathcal{J}}(\rho_{AB})$ has an operational interpretation as a measure of classical correlations, as the distillable common randomness (DCR) with one-way classical communication\cite{dw04}, which is identical to the regularized version of the measure of classical correlations as defined by Henderson and Vedral\cite{henderson01a}. Using the monogamy between DCR and the entanglement cost $E_C,$ the regularized version of the entanglement of formation~\cite{kw04,hhhh07}, it has been shown that quantum discord is subadditive\cite{md10}. Thus, the operational and other interpretations of quantum discord based on multicopy quantum protocols only provide a lower bound. Interestingly, for separable states, quantum discord is additive. This follows from the trivial additivity of $E_C$ for separable states.

\section{The mother protocol and the quantum information family tree}

Abeyesinghe \textit{et al.} showed that essentially all unidirectional, bipartite and memoryless quantum communication protocols are actually siblings originating from one ``mother". The mother protocol can be seen to provide a hierarchical structure to the family of quantum protocols\cite{adhw09}.

The mother protocol starts with $n$ copies of a quantum state $ \ket{\psi^{ABR}}$.  Alice holds the $A$ shares and Bob the $B$ shares. The reference system $R$ is ``purification" of the $AB$ system (which might be described by a mixed state) and does not actively participate in the protocol. The mother protocol can be viewed as an entanglement distillation between $A$ and $B$ when the only type of communication permitted is the ability to send qubits from Alice to Bob. The transformation can be expressed in the resource inequality as
\be
\label{mother}
\langle\psi^{AB}\rangle  + \frac{1}{2} I (A:R) [q \rightarrow q] \geq \frac{1}{2} I (A:B)[qq].
\ee
Here, $\ket{\psi^{AB}}$ refers to the state shared between Alice and Bob whose purification is the state $\ket{\psi^{ABR}}$. The above inequality states that $n$ copies of the state $|\psi^{AB}\rangle$ can be converted to $\frac{1}{2} I (A:B)$ EPR pairs per copy, provided Alice is allowed to communicate with Bob by sending him qubits at the rate $\frac{1}{2} I (A:R)$ per copy.

A stronger version of the mother protocol, the FQSW protocol not only enables the two parties, Alice ($A$) and Bob ($B$), to distill  $\frac{1}{2} I (A:B)$ EPR pairs per copy, in addition Alice can ``merge" her state with Bob. This implies that Alice is able to successfully transfer her entanglement with the reference system $R$ to Bob. Writing the FQSW in terms of a resource inequality
\ben
\langle\psi^{AB}\rangle  &+& \frac{1}{2} I (A:R) [q \rightarrow q] \geq \frac{1}{2} I (A:B)[qq] \nonumber \\
    && +  \mbox{State Merging between}~A~\mbox{and}~ B
\een
In a more rigorous mathematical notation, we write the above as
\ben
 \label{fqsw}
\langle \mathcal{U}^{S \rightarrow AB} : \psi^{S}\rangle  + \frac{1}{2} I (A:R) [q \rightarrow q] &\geq& \frac{1}{2} I (A:B)[qq] \nonumber \\
    &+& \langle \mathbb{I}^{S \rightarrow \hat{B}} : \psi^{S}\rangle,
\een
where we have a noisy resource mixed state, $ \psi^{S}$ inserted between a ``$\langle \cdot  \rangle$". Thus a mixed state is represented by $\langle \psi^{S} \rangle $, and a noisy channel by $\langle \mathcal{N}\rangle$. A channel is a relative resource $\langle \mathcal{U}^{S \rightarrow AB} : \psi^{S}\rangle$ meaning that the protocol only works provided the input to the channel is the state $\psi^{S}$. On the LHS, $\mathcal{U}$ takes the state $ \psi^{S}$ and distributes it to Alice and Bob. On the RHS, the symbol $\mathbb{I}$ is an identity channel taking the state $\psi^S$ to Bob alone. The state $\psi^{S}$ on the left-hand side of the inequality is distributed to Alice and Bob, while on the right-hand side, that same state is given to Bob alone. This inequality states that starting from the state  $|(\psi^{ABR})^{\otimes n}\rangle$, and using $\frac{1}{2} I (A:R)$ bits of quantum communication from Alice to Bob, they can distill $\frac{1}{2} I (A:B)$ EPR pairs per copy, and in addition Alice can accomplish merging her state with Bob, in which she is able to successfully transfer her entanglement with the reference system $R$ to Bob. This means that Alice transfers her portion of the state to Bob. In other words, they manage to create the state $|(\psi^{R\hat{B}})^{\otimes n}\rangle$, where $\hat{B}$ is a register held with $B$ and  $|(\psi^{R\hat{B}})\rangle = |(\psi^{ABR})\rangle$ in the limit $n \rightarrow \infty$. Finally, the asymptotic nature of the equivalence is denoted by the symbol $\geq$.

\section{Quantum discord as a measure of coherence in the FQSW protocol}
\label{sec:Main}

The FQSW is essentially a non-dissipative protocol in that no information is leaked to the environment in each step of the protocol, but any practical implementation of a quantum information protocol will be affected by loss and noise. In particular, we will consider loss of information and coherence at Bob's end. This can be studied by considering a unitary coupling between Bob's system $B$ and an ancillary environment system, say $C,$ and then tracing $C$ out. Physically, such a quantum operation will emulate environmental decoherence.

We begin by expanding the size of the Hilbert space so that an arbitrary measurement (or any other quantum operation) can be modeled by coupling to the auxiliary subsystem and then discarding it. We assume the ancilla $C$ to initially be in a pure state $\ket{{0}}$, and a unitary interaction $U$ between $B$ and $C$. Letting primes denote the state of the system after $U$ has acted we have $S(A,B) = S(A',B'C')$ as $C$ starts out in a product state with $AB$. We also have $I(A : BC) = I(A' : B'C')$. As discarding quantum systems cannot increase the mutual information, we get $I(A' : B') \leq I(A' : B'C')$.

Now consider the FQSW protocol between $A$ and $B$ in the presence of $C$. We can always view the yield of the FQSW protocol on the system $AB$ to be the same as that of performing the protocol between systems $A$ and $BC$, where $C$ is some ancilla (initially in a pure state) with which $B$ interacts coherently through a unitary $U$. Such an operation does not change the cost or yield of the FQSW protocol, as shown, but helps us in counting resources. Discarding system $C$ yields
\be
\label{eq:Idiff1}
I(A' : B') \leq I(A':B'C') = I(A:BC) = I(A : B),
\ee
or alternatively,
\be
\label{eq:Sdiff}
S(A'|B') \geq S(A'|B'C') = S(A|B).
\ee
Now consider a protocol which we call as $FQSWD_{B}$  (FQSW after decoherence), where the subscript refers to the decoherence at $B$. The resource inequality for $FQSWD_{B}$ is
\ben
\avg{\mathcal{U}^{S' \rightarrow A'B'} : \psi^{S'}}  + \frac{1}{2} I (A':R') [q \rightarrow q] &\geq& \frac{1}{2} I (A':B')[qq] \nonumber \\
    && + \avg{\mathbb{I}^{S' \rightarrow \hat{B}} : \psi^{S'}}. \nonumber
\een
The primed letters, $A'$, $B'$ etc., indicate that the protocol is taking place in the presence of decoherence at Bob's end.

As in the fully coherent version, Alice is able to transfer her entanglement with the reference system $R'$, and is able to distill $\frac{1}{2} I (A':B')$ EPR pairs ($[qq]$) with Bob. The net quantum gain for the fully coherent protocol is $G=\frac{1}{2} I (A:B) -\frac{1}{2} I (A:R) = -S(A|B)$ ebits. This is the difference between the yield obtained and the cost of quantum communication incurred. Likewise, the net gain for the protocol suffering decoherence at $B$ is $G_{D}=\frac{1}{2}I(A':B') - \frac{1}{2}I(A':R') = -S(A'|B')$. Therefore, the net advantage of the coherent protocol over the decohered one is given by  $D = G-G_{D} = S(A'|B') - S(A|B)$ ebits. Evidently, this quantum advantage depends on the exact nature of the environment and the system's interaction with it via $U.$ Employing the original definition of quantum discord due to Zurek\cite{z00} (Zurek's original definition of discord did not consider optimizing over all measurements), $D$ quantifies the loss in the yield of a quantum protocol due to environmental decoherence. Our results therefore provide a standard way of quantifying, in entropy units, the damage to the performance of quantum process and protocols in the presence of any decoherence process in any experimental scenario.

The strength of our result, however, comes from the next step of minimizing $D$ over all environmental operations performing measurements. Using the measurement model of quantum operations\cite{nc00}, the state $\rho_{AB}$ under measurement of subsystem $B$, changes to $\rho_{AB}' = \sum_{j} p_{j} \rho_{A|j} \otimes \pi_j,$ where $ \{\pi_j\} $ are orthogonal projectors resulting from a Neumark extension of the POVM elements\cite{peres}. The unconditioned post-measurement states of $A$ and $B$ are
\be
\label{eq:cond}
\rho_{A}' =\sum_{j} p_{j} \rho_{A|j} = \rho_{A},~~~\rho_{B}' = \sum_{j} p_{j} \pi_j.
\ee
Invoking these relations, we get
\be
S(A'|B') = \sum_{j} p_{j} S(\rho_{A|j}).
\ee
After minimization over all POVMs, $D$ reduces to $\mathcal{D}(A:B)$ as defined in Eq.~(\ref{discexp}). \textit{Quantum discord thus quantifies the minimum loss in yield of the FQSW protocol due to decoherence.} This is our main result, and shows that the performance of all the protocols in the quantum information family tree must be judged by the quantum discord. The connection between quantum discord and the FQSW protocol provides a metric for studying the advantage of coherence in accomplishing any of the children protocols that can be derived from the FQSW protocol. For example, we look at the noisy versions of quantum teleportation, super-dense coding, and entanglement distillation.

The connection we have made here is crucial. While it is known that entanglement is necessary for the success of the protocol, and that the presence of decoherence affects its performance, we have now provided a quantitative result of the amount of such a depreciation. We have shown that the amount by which a protocol suffers in the presence of decoherence is an inherent property of the quantum states involved. It suggests that the best state to be employed in a noisy quantum communication protocol should be the outcome of a tradeoff between the entanglement and discord of the state involved, since the variation of discord and entanglement in quantum states in not monotonic\cite{modi10,alqasimi11,girolami11}. In the next section,  we demonstrate the versatility of this result by applying it to some well-known quantum protocols.

\section{Quantum discord in the children protocols}
\label{sec:child}
The connection between quantum discord and the FQSW protocol provides a metric for studying the effect of coherence in accomplishing any of the so called ``children protocols" that can be derived from the FQSW protocol. In this section, we show that by connecting quantum discord with the FQSW protocol, we can interpret discord as the advantage of quantum coherence in noisy versions of teleportation, super-dense coding, and entanglement distillation. Finally, we reproduce an earlier result on the connection of quantum discord and quantum state merging.

\subsection{Noisy teleportation}

The noisy teleportation resource inequality can be expressed as
\be
\label{nt}
\langle\Psi^{AB}\rangle  + I (A:B) [c \rightarrow c] \geq I (A\rangle B)[q \rightarrow q],
\ee
obtained by combining the mother protocol with teleportation\cite{dhw08}.  Here, $I(A\rangle B) = -S(A|B)$ is also known as the coherent information\cite{sn96}. When Bob undergoes decoherence, we get,
\begin{equation}
\label{ntd}
\langle\Psi^{A'B'}\rangle  + I (A':B') [c \rightarrow c] \geq I (A'\rangle B')[q \rightarrow q].
\end{equation}
The above can be interpreted as following: The net loss in the number of qubits that can be teleported when comparing the coherent teleportation (the one without any decoherence), Eq.~(\ref{nt}), and the one which suffers decoherence, Eq.~(\ref{ntd}), is given by  $I(A\rangle B) - I(A'\rangle B') = S(A'|B') - S(A|B)$. We assume the classical communication to be free in this case, as long as we are teleporting unknown quantum states. We have $S(A|B) = S(A) - I(A : B) = S(A) - I(A : BC) = S(A|BC).$ As in Sec.~(\ref{sec:Main}), the application of the unitary $U$, but before discarding the subsystem $C$, the cost of teleportation is still given by $S(A'|B'C') = S(A|B)$. From Eq.~(\ref{eq:Idiff1}),
\be
\label{eq:Sdiff1}
S(A'|B') \geq S(A'|B'C') = S(A|B).
\ee
Therefore, we see that the advantage of the coherent protocol over the noisy version in teleporting unknown quantum states is equal to the quantum discord of the original state.

For a particular class of two-qubit quantum states, it was recently shown that the fidelity of remote state preparation is equal to the geometric quantum discord\cite{dvb10}. This was also demonstrated experimentally using photonic qubits\cite{dakic12}. Remote state preparation is a special case of quantum teleportation, and the relation between discord and the fidelity has been suggested as an operational interpretation for quantum discord.

\subsection{Noisy super-dense coding}

Noisy super-dense coding can be derived by combining the mother protocol with super dense coding\cite{dhw08}
\be
\label{superdense}
 [qq]  +  [q \rightarrow q] \succeq  2[c \rightarrow c],
\ee
showing that one can employ a shared ebit and a single bit of quantum communication to communicate 2 bits of classical information. Here, $[q \rightarrow q]$ represents one qubit of communication between two parties and $[qq]$ represents one shared ebit between two parties. Similarly, $[c \rightarrow c]$ represents one classical bit of communication between the parties.
The symbol $\succeq $ is used to denote exact attainability as compared to $\geq$ which is to denote asymptotic attainability. Combining these, the noisy super-dense coding protocol can be expressed as,
\be
\langle\Psi^{AB}\rangle  + S(A) [q \rightarrow q] \geq I (A:B)[c \rightarrow c].
\ee
When the party $B$ is undergoing decoherence, the noisy super-dense coding can be expressed as,
\be
\langle\Psi^{A'B'}\rangle  + S(A') [q \rightarrow q] \geq I (A':B')[c \rightarrow c].
\ee
We note that $S(A) =S(A')$. Thus, due to decoherence, the number of classical bits communicated through this protocol gets reduced by the amount
$I (A:B) -  I (A':B')$, which is equal to the discord of the original state.

While all our results are derived for finite-dimensional cases, gaussian quantum discord\cite{ad10} has been related to a generalisation of quantum dense coding for continuous-variable states, when all the states and operations involved are gaussian. The problem was cast as the advantage that can be harnessed by using nonlocal quantum interactions. This connection was also explored experimentally in the same work\cite{gu12}.

\subsection{Entanglement distillation}

The one-way entanglement distillation can be expressed as
\be
\langle\Psi^{AB}\rangle  + I (A:R) [c \rightarrow c] \geq I (A\rangle B)[qq].
\ee
This inequality can be derived by combining the FQSW protocol Eq.~(\ref{fqsw}) and recycling the $\frac{1}{2} I(A:R)$ ebits out of the total $\frac{1}{2} I(A:B)$ produced for teleportation, as shown in\cite{dhw08}. Decoherence at Bob's end $B$ provides
\be
\langle\Psi^{A'B'}\rangle  + I (A':R') [c \rightarrow c] \geq I (A'\rangle B')[qq].
\ee
The net change in entanglement distillation is $I(A' \rangle B') - I(A\rangle B) = S(A|B) - S(A'|B'),$ which is the negative of the quantum discord of the original state. As is well known, classical communication between parties cannot enhance entanglement, and we can neglect the overhead of $I(A:R)-I(A':R')$ classical bits.

\subsection{Quantum state merging}

Quantum state merging protocol is the extension of the classical Slepian-Wolf protocol into the quantum domain where Alice and Bob share the quantum state $\rho_{AB}^{\otimes n}$, with each party having the marginal density operators $\rho_{A}^{\otimes n}$ and $\rho_{B}^{\otimes n}$ respectively. Let $\ket{\Psi_{ABC}}$ be a purification of $\rho_{AB}.$ Assume, without loss of generality, that Bob holds $C.$ The quantum state merging protocol quantifies the minimum amount of quantum information which Alice must send to Bob so that he ends up with a state arbitrarily close to $\ket{\Psi}_{B'BC}^{\otimes n},$ $B'$ being a register at Bob's end to store the qubits received from Alice. It was shown that in the limit of $n\rightarrow \infty$, and asymptotically vanishing errors, the answer is given by the quantum conditional entropy: $S(A|B) = S(A,B) - S(B)$. When $S(A|B)$ is negative, Bob obtains the full state with just local operations and classical communication, and distill $-S(A|B)$ ebits with Alice, which can be used to transfer additional quantum information in the future. Quantum state merging provides an operational interpretation for quantum discord\cite{md10,cabmpw10}. It is the markup in the cost of quantum communication in the process of quantum state merging, if one discards relevant prior information. An intuitive argument for the above interpretation of quantum discord can be made through strong subadditivity\cite{how07}
\be
\label{eq:ssa}
S(A|B,C) \leq S(A|B).
\ee
From the point of view of the state merging protocol, this has a very clear interpretation. Having more prior information makes state merging cheaper. In other words, throwing away information will make state merging more expensive. Thus, if Bob discards system $C$, it will increase the cost of quantum communication needed by Alice in order to merge her state with Bob.

Quantum state merging can be derived from the FQSW if the entanglement produced at the end of the FQSW protocol can be used to perform teleportation. As a resource inequality
\be
\label{sm}
\avg{\Psi^{AB}}  + S(A|B) [q \rightarrow q] + I (A:B)_{\Psi} [c \rightarrow c] \geq \avg{\mathbb{I}^{S \rightarrow \hat{B}} : \Psi^{S}},
\ee
it accomplishes state merging from Alice to Bob at the cost of $S( A|B)$ bits of quantum communication. When $S(A|B)$ is negative, Alice and Bob can distill this amount of entanglement in the form of Bell pairs. Thus, quantum state merging provides an operational interpretation of $S(A|B).$ As in Sec.~(\ref{sec:Main}), the resource inequality for the noisy version of the quantum state merging protocol
\be
\avg{\Psi^{A'B'}}  + S(A'|B') [q \rightarrow q] + I (A':B')_{\Psi} [c \rightarrow c] \geq \avg{\mathbb{I}^{S \rightarrow \hat{B}} : \Psi^{S}}.
\ee
The cost of quantum communication in this case is $S(A'|B'),$ and the mark up in this cost is $S(A'|B') - S(A|B)$, which is equal to the quantum discord of the original state.

\section{Summary and Outlook}

The role of quantum entanglement as a resource in quantum information science is well established and acknowledged. It is also well known that maximizing the amount of entanglement in a system does not monotonically enhance the quantum advantages it may provide. This is true at a conceptual level due to results such as the Gottesman-Knill theorem\cite{nc00}, as well as at a practical level, where the more entangled the state, the more it is susceptible to decoherence. Our work clarifies and establishes the central role played by quantum discord in the latter scenario. We have shown that quantum discord is a quantity of fundamental significance in quantum information theory, by virtue of its role in the performance of a large class of quantum communication protocols.

To harness the enhancements promised by quantum technologies in the real world, it is essential that we go beyond the idealized scenarios in which most of the now-famous protocols such as teleportation, and dense-coding were designed. This article summarizes the recent advances made in that direction. The outcome is that quantum discord quantifies, in a very direct manner, the damage that a decohering environment inflicts on the advantages promised by a quantum protocol. If our goal is to maximize the extraction of such quantum advantages, we must design quantum states that minimize the deleterious effects of the environment. This inexorably leads us to identifying quantum states that provide a balance between its entanglement and discord content. Given the nontrivial interrelation between the two quantities, and the geometry of entangled and discordant states, this provides a promising and engaging avenue for future research.

The generality of the framework -- that of the FQSW protocol -- employed by us also allows for additional scopes of progress. It would be fruitful to extend the protocol to include multiple parties and multiple rounds of communications, and then explore the role of quantum discord in the advantages attainable in such scenarios. Broader applicability would also result from the incorporation of non-markovian environments into the framework. Activities in this direction are already being undertaken\cite{vasile10,haikka12}, and unified framework will put such results in context. It will also allow for the theoretical and experimental exploration of quantum advantages provided by discord in atomic, molecular and condensed systems.

\section*{Acknowledgements}
It is a pleasure to that several discussions with G. Adesso, C. Brukner, B. Dakic, M. Gu, K. Modi, M. Piani, and A. Shaji. VM was supported by the Center for Quantum Information and Control (CQuIC), Ontario Ministry of Economic Development and Innovation, NSERC, and the Laurier Research Office. AD was supported in part by the EPSRC (Grant Nos. EP/H03031X/1 and RDF/BtG/0612b/31) and the EU~Integrated Project QESSENCE.

\section*{References}

\bibliography{DiscordRefs}

\end{document}